\documentclass[a4paper,12pt]{article}
\usepackage[pctex32]{graphics}

\begin{document}

\topmargin 0pt
\oddsidemargin 0mm

\begin{titlepage}
\begin{flushright}
{\tt gr-qc/0702145}
\end{flushright}

\vspace{5mm}
\begin{center}
{\Large \bf Quantum Cooling Evaporation Process \\ in Regular
Black Holes } \vspace{12mm}

{\large   Yun Soo Myung$^{\rm a,}$\footnote{e-mail
 address: ysmyung@inje.ac.kr},
 Yong-Wan Kim $^{\rm a,}$\footnote{e-mail
 address: ywkim65@gmail.com},
and Young-Jai Park$^{\rm b,}$\footnote{e-mail
 address: yjpark@sogang.ac.kr}}
 \\
\vspace{10mm} {\em $^{\rm a}$Institute of Mathematical Science and
School of Computer Aided Science, \\Inje University, Gimhae 621-749,
Korea \\} {\em $^{\rm b}$Department of Physics  and Center
for Quantum Spacetime,\\ Sogang University, Seoul 121-742, Korea}
\end{center}

\vspace{5mm}

\centerline{{\bf{Abstract}}}

\vspace{5mm} We investigate a universal behavior of thermodynamics
and evaporation process for  the regular black holes. We newly
observe an important point where the temperature is maximum, the
heat capacity is changed from negative infinity to positive
infinity, and the free energy is minimum. Furthermore, this point
separates the evaporation process into the early stage with
negative heat capacity and  the late stage with positive heat
capacity. The latter represents the quantum cooling evaporation
process. As a result, the whole evaporation process could be
regarded as the inverse Hawking-Page phase transition.

\vspace{3mm}

\end{titlepage}

\newpage

\renewcommand{\thefootnote}{\arabic{footnote}}
\setcounter{footnote}{0} \setcounter{page}{2}

{\it Introduction.---}Hawking's semiclassical analysis of  the
black hole radiation suggests that most  information about initial
states is shielded behind the event horizon and will not back to
the asymptotic region far from the evaporating black
hole~\cite{HAW1}. This means that the unitarity is violated by an
evaporating black hole. However, this conclusion has been debated
by many authors for three decades~\cite{THOO,SUS,PAG}. It is
closely related  to a long standing puzzle of  the information
loss paradox, which states the question of whether the formation
and subsequent evaporation
 of a black hole is unitary. One of the most urgent problems in black
hole physics is  to resolve the unitarity issue. In this direction
a complete description of black hole evaporation is an important
issue. In order to determine the final state of evaporation
process, a more precise treatment including
 quantum gravity effects  and backreaction is generally required. At present, two
leading candidates for quantum gravity are the string theory and
the loop quantum gravity. Interestingly, the semiclassical
analysis of the loop quantum black hole provides a regular black
hole (RBH) without singularity  in contrast to the classical
one~\cite{MOD}. Its minimum size $r_c$ is at Planck scale
$\ell_{Pl}$. On the other hand, in the continuing search for
quantum gravity, the black hole thermodynamics may be related to a
future experimental result at the LHC~\cite{LHC}.

RBHs have been considered, dating back to Bardeen~\cite{BAR}, for
avoiding the curvature singularity beyond the event horizon in
black hole physics~\cite{RBH}. Their causal structures are similar
to the Reissner-Nordstr\"{o}m (RN) black hole with the singularity
replaced by  de Sitter space-time with curvature radius
$r_0=\sqrt{3/\Lambda}$~\cite{Dymn}. Recently, Hayward has
discussed the formation and evaporation process of a RBH with
minimum size $l$~\cite{HAY}, which can be identified with the
minimal length induced  from the string theory~\cite{Vene}. A more
rigorous treatment of the evaporation process was carried out for
the renormalization group (RG) improved black hole with minimum
size $r_{cr}=\sqrt{\tilde{\omega}G}$~\cite{BR1,REU}. The
noncommutativity also provides another RBH with minimum scale
$\sqrt{\theta}$: noncommutative black hole~\cite{SS}. Very
recently, we have investigated thermodynamics and evaporation
process of the noncommutative black hole~\cite{YKP}. The RN black
hole with charge $Q$ also belongs to the RBH~\cite{Hisc}, even
though it has a timelike singularity~\cite{RNBH}. It turned out
that the final state of the evaporation process for all RBHs is a
cold, Planck size remnant of the extremal black hole. The
connection between their minimum sizes is given by $r_c \sim r_{0}
\sim l \sim r_{cr} \sim \sqrt{\theta} \sim Q \sim \ell_{Pl}$.

It is very important to study the terminal phase of  black hole
evaporation. In the semiclassical study of the Schwarzschild black
hole, the temperature ($T_H \propto 1/m$) and the luminosity
($L_{Sch} \propto 1/m^2$) diverge as $m$ approaches zero. This
means that the semiclassical approach breaks down for very light
holes. Furthermore, one has to take into account the backreaction.
It was shown that the effect of quantum gravity could cure this
pathological short distance behavior~\cite{nicol}.

In this Letter, we first study universal thermodynamic properties
of RBHs by  analyzing the minimal model proposed by
Hayward~\cite{HAY} and then investigate its evaporation process.
We wish to remind the reader that the RBH  is closely related to
effects of quantum  gravity.  Especially, we newly observe an
important point at $r_+=r_m$ where the temperature is maximum, the
heat capacity is changed from negative infinity to positive
infinity, and the free energy is minimum. This point separates the
whole evaporation process into the early stage with negative heat
capacity and  the late stage with positive heat capacity. The
latter is described by the quantum cooling evaporation process
(QCEP) which is a necessary step to reach extremal black hole. For
the QCEP, the temperature decreases near Planck scale as the mass
of black hole decreases, while for the early evaporation process,
the temperature increases as the mass of  black hole decreases. It
is important to note that  we do not need to take into account the
backreaction for RBHs due to the QCEP.

We could understand the thermodynamic process for RBHs from the
analogy of the Hawking-Page (HP) phase transition in the AdS black
hole~\cite{Hawking1, Hawking2}. The relevant thermodynamic
quantities are temperature $T_{SADS}$, heat capacity $C_{SADS}$,
free energy $F_{SADS}$ and, off-shell free energy
$F^{off}_{SADS}$\footnote{Their explicit forms are
$T_{SADS}=\frac{1}{4\pi}\Big(\frac{3r_+}{\ell^2}+\frac{1}{r_+}\Big),~C_{SADS}=2
\pi
r_+^2\frac{(3r_+^2+\ell^2)^2}{(3r_+^2-\ell^2)^2},
~F_{SADS}=\frac{r_+}{4}\Big(1-\frac{r_+^2}{\ell^2}\Big)$,~
$F^{off}_{SADS}=\frac{r_+}{2}\Big(1+\frac{r_+^2}{\ell^2}\Big)-\pi
r_+^2 T$ with the curvature radius $\ell$ of AdS$_4$ spacetime.
Here we have $r_0=\frac{\ell}{\sqrt{3}} $ where the heat capacity
blows up and the temperature has the minimum value
$T_{0}=\frac{\sqrt{3}}{2\pi \ell}$ and $r_1=\ell $ where the free
energy is zero and the temperature has the critical value
$T_{1}=\frac{1}{\pi \ell}$. For numerical computations, we choose
$\ell=10$.}. In the HP transition, one generally starts with thermal radiation
in AdS space. A small black hole appears with negative heat
capacity. The heat capacity changes from negative infinity to
positive infinity at the minimum temperature $T_0$. Finally, the
large black hole with positive heat capacity comes out as a stable
object. There is a change of the dominance at the critical
temperature $T_1$: from thermal radiation to black hole.

In contrast to the HP case, we start with the large unstable black hole
with negative heat capacity for RBHs. The heat capacity changes
from negative infinity to  positive infinity at the maximum
temperature. Then, the small black hole with positive heat
capacity comes out. There is a change of the dominance at the
critical temperature near $T=0$: from  a large  black hole to a
different, extremal black hole. Consequently, we regard the
evaporation process of RBHs as the inverse HP transition because
this is the process from initial (unstable) large black hole to
final (stable) extremal black hole. We note that the QCEP plays a
crucial role in the inverse HP transition. However, it takes an
infinite time to reach the final remnant of extremal black hole
using the quantum-corrected Vaidya metric.

{\it Thermodynamics of regular black holes.---}It was shown that
in order to obtain a RBH, we need to introduce an anisotropic
fluid whose energy-momentum tensor is given by $T^\mu_{~~\nu}={\rm
diag}[-\rho,p_r,p_\bot,p_\bot]$ with  energy density $\rho$,
radial pressure $p_r$, and tangential pressure $p_\bot$. For
simplicity,  we study the minimal model~\cite{HAY} provided by the
energy-momentum tensor
\begin{equation} \label{2eq1}
\rho=\frac{3 l^2m^2}{2 \pi(r^3+2l^2m)^2}=-p_r,~~
p_\bot=\frac{3l^2m^2(r^3-l^2m)}{\pi(r^3+2l^2m)^3}
\end{equation}
with the Planck units of $c=\hbar=G=\ell_{Pl}=1$.  Solving the
Einstein equation $G_{\mu\nu}=8 \pi T_{\mu\nu}$ leads to the
solution

\begin{equation} \label{2eq2}
ds^2_{RBH} \equiv g_{\mu\nu} dx^\mu dx^\nu= -F(r)dt^2+
F(r)^{-1}dr^2+r^2d\Omega^2_2.
\end{equation}
The metric function $F(r)$ is given by

\begin{equation} \label{2eq3}
F(r)=1-\frac{2mr^2}{r^3+2l^2m},
\end{equation}
where $l$ denotes the curvature radius of de Sitter space-time
near the center and $m=4\pi\int^{\infty}_{0}\rho(r)r^2 dr$
represents the Arnowitt-Deser-Misner mass. We have de Sitter
space-time $F(r) \simeq 1-r^2/l^2$ as $r \to 0$, while an
asymptotically Schwarzschild space-time $F(r) \simeq 1-2m/r$
appears as $r \to \infty$. Hence,  $\rho$ connects the de Sitter
vacuum in the origin with the Minkowski vacuum at infinity.

 From the condition of horizon $F=0$, we obtain the horizon
 masses  $m_{\pm}=r^3_{\pm}/2(r^2_{\pm}-l^2)$.
Here we find  the minimum mass $m_*=3\sqrt{3}l/4$ at
$r_*=\sqrt{3}l$. For definiteness,  we consider three different
types: i) For $m>m_*$, two distinct horizons appear with the inner
cosmological horizon $r=r_-(l < r_- \le r_*)$ and the outer event
horizon $r=r_+(r_* \le r_+ <\infty)$. They are analytically derived
by $r_+=\frac{m}{3}(2+ 4 \cos\frac{\alpha}{3}),~
r_-=\frac{m}{3}(2+ 4 \cos(\frac{\alpha}{3}-\frac{2\pi}{3}))$ where
$\cos \alpha=1-\frac{2m_*^2}{m^2}$ with $\frac{2m_*}{m} < \alpha
\le \pi$. In particular, for  $m \gg m_*(\alpha \to
\frac{2m_*}{m}\simeq 0)$,
 the outer horizon is located at $r_+\simeq 2m$,
 while the inner horizon is at $r_-\simeq l$. ii)
In case of $m=m_*(\alpha = \pi)$, one has a degenerate horizon at
$r=r_*$, which corresponds to the extremal black hole. iii) For
$m<m_*$, there is no horizon.

The black hole temperature  can be calculated
 to be \begin{equation} \label{2eq4}
T_{RBH}(r_+)=
\frac{1}{4\pi}~\Bigg[\frac{dF}{dr}\Bigg]_{r=r_+}=\frac{1}{4\pi
r_+}\frac{r_+^2-r_*^2}{r_+^2}
\end{equation}
with a fixed core radius $l=1$.  For $r_+^2\gg 1$, one recovers
the Hawking temperature $T_H=1/4\pi r_+$ of the Schwarzschild
black hole. Therefore, at the early stage of the Hawking
radiation, the black hole temperature increases as  the horizon
radius decreases. It is important to investigate what happens as
$r_+ \to 0$. In the Schwarzschild case, $T_H$ diverges and this
puts the limit on the validity of the evaporation process via the
Hawking radiation. Against this scenario, the temperature
$T_{RBH}$ includes quantum effects, which are relevant at short
distance comparable to the Planck scale of $r_+ \simeq
1$~\cite{BR1,SS}.
 As is shown in Fig.1, the temperature of the RBH grows
until it reaches to the maximum value $T_m=0.017$ at
$r_+=r_m=3(m=m_m=1.68)$ and then falls down to zero at
$r_+=r_*=\sqrt{3}(m=m_*)$ which the extremal black hole appears
with $T_{*}=0$. As a result, the evaporation  process is split
into the right branch of $r_m < r_+< \infty$ called the early
stage of evaporation and the left branch of $ r_\ast < r_+<r_m$
called the QCEP. In the region of $r<r_*$, there is no black hole
for $m<m_*$ and thus the temperature cannot be defined. For
$m>m_*$, we have the inner horizon at $r=r_-$ inside the outer
horizon but an observer at infinity does not recognize the
presence of this cosmological horizon. Hence, we regard this
region as the forbidden region.
\begin{figure}[t!]
   \centering
   \includegraphics{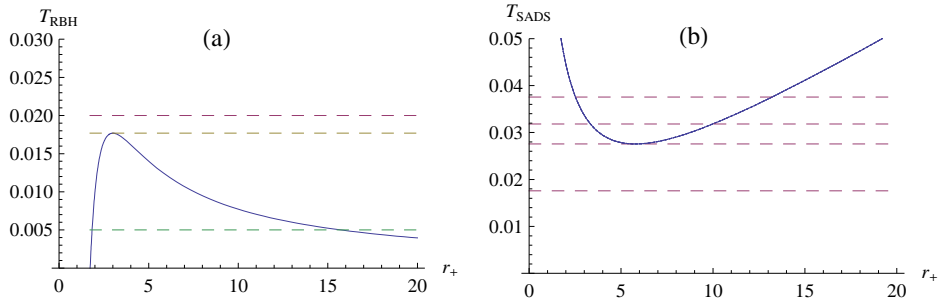}
\caption{(a) The solid line represents  temperature $T_{RBH}$ with
the maximum point at $r_+=r_m$ and minimum point at $r_+=r_\ast$.
The near-horizon region where the QCEP takes place is  $r_\ast <
r_+ <r_m$. Three horizontal dashed lines denote the temperature
$T=0.02$, $T_m$ and $0.005$ from the top to the bottom. (b)
Temperature for Schwarzschild-AdS black hole with the minimum
point at $r_+=r_0$. } \label{fig1}
\end{figure}

The entropy $S_{RBH}=\int_{r_*}^{r_+}(m^\prime/T_{RBH})dr$ of the
RBH can be obtained from the first-law of thermodynamics
$dm=T_{RBH}dS_{RBH}$ as
\begin{equation} \label{2eq5}
\label{entropy} S_{RBH}(r_+)=\frac{A}{4}+
\frac{\pi}{50}\Bigg[-\frac{10}{2r_+^2-1}+108 \ln(r_+^2-r_*^2)+17
\ln(2 r_+^2-1) \Bigg]
\end{equation}
with the area of the event horizon $A=4 \pi r_+^2$.   We have
negative infinity-entropy for the extremal black hole at $r_+=r_*$
due to the third term. Hence we cannot find logarithmic correction
to the extremal black hole. On the other hand, we have the
area-law behavior of $S_{BH} \simeq \pi r_+^2$ for $r_+ \gg1$.
\begin{figure}[t!]
   \centering
   \includegraphics{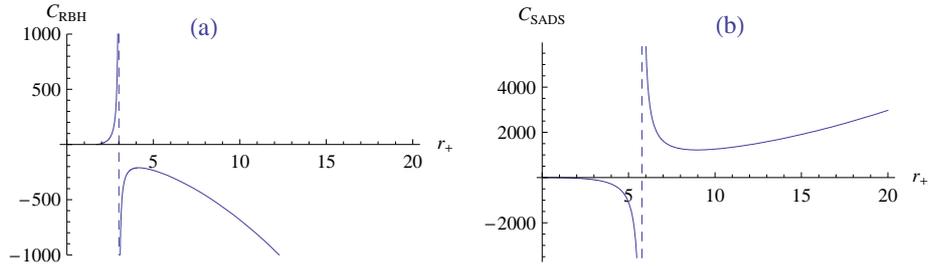}
\caption{(a) The solid line represents heat capacity $C_{RBH}$ as
a function of the black hole radius $r_+$. The QCEP takes place in
the near-horizon region of  $r_* < r_+ <r_m(C \ge 0)$. (b) The
heat capacity of Schwarzschild-AdS black hole is negative for
$r_+<r_0$ and  positive value for  $r_+>r_0$. } \label{fig2}
\end{figure}

In order to check the thermal stability of the RBH, we have to
know the heat capacity~\cite{BR1}.  Its heat capacity
$C_{RBH}=dm/dT_{RBH}$ is given by
\begin{equation} \label{2eq6}
C_{RBH}(r_+)=- 2 \pi r_+^2~\frac{
r_+^4(r_+^2-r_*^2)}{(r_+^2-1)^2(r_+^2-r_m^2)}
\end{equation}
and its variation  is plotted in Fig. 2. Here, we find the
near-horizon region of $C_{RBH}>0$, where the QCEP takes place.
This means that the RBH could be thermodynamically stable in the
range of $r_*< r_+ <r_m$. The heat capacity becomes singular at
$r_+=r_m$, which corresponds to the maximum temperature $T=T_m$. We
also observe that a thermodynamically unstable region ($C_{RBH}<
0$) appears for $r_+>r_m$.  We note that in the Hawking regime of
$r_+ \gg 1$, $C_{RBH}$ is consistent with the specific heat
$C_{RBH} \simeq -2\pi r^2_+$ of the Schwarzschild black hole. Also
we have $C_{RBH}|_{r_+=r_*}=0$ for the extremal black hole.

Now, we are in a position to discuss a possible phase transition.
For this purpose, we introduce the on-shell free energy as
\begin{equation} \label{2eq7}
F_{RBH}(r_+)=m(r_+)-m_*-T_{RBH}(r_+)S_{RBH}(r_+),
\end{equation}
where for fixed $l=1$, we have to use the extremal black hole as
background~\cite{CEJM}.  Its graph is shown in Fig. 3.
Interestingly, the free energy has the minimum value at $r_+=r_m$.
The QCEP takes place for $r_* < r_+ <r_m$. For $r_+ \gg 1$, one
recovers $F_{RBH}\simeq r_+/4$ for the Schwarzschild black hole.
Further, one needs to know the off-shell free-energy
\begin{equation} \label{2eq8}
F^{off}_{RBH}(r_+,T)=m(r_+)-m_*-T S_{RBH}(r_+)
\end{equation}
with the temperature  $T$ of the heat reservoir.
\begin{figure}[t!]
   \centering
   \includegraphics{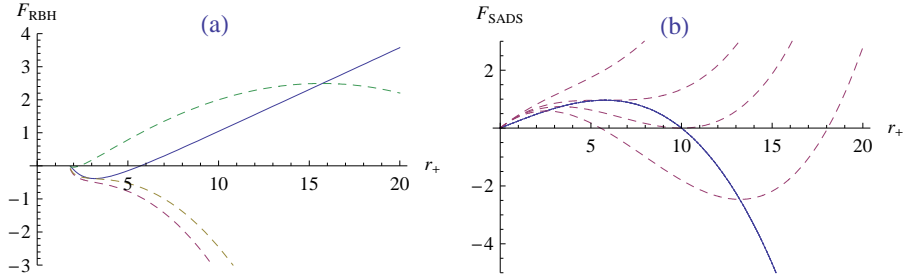}
   \caption{(a) The solid line represents plot of the  free energy $F_{RBH}$ as a function of $r_+$.
   The dashed curves denote $F^{off}_{RBH}(r_+,T=0.005),~
   F^{off}_{RBH}(r_+,T=T_m),~F^{off}_{RBH}(r_+,T=0.02)$ from top to bottom.
   $r_i=15.72$ and $r_*=1.84$ represent the starting  point and
 the ending point for an evaporation process at $T=0.005$, respectively. (b) The free energy and off-shell
 free energy for are shown for the Schwarzschild-AdS black hole. We find the HP transition
 along the bottom dashed curve: starting with thermal radiation at $r_+=0$ and ending with large stable black hole
 at $r_+=r_s$.} \label{fig3}
\end{figure}

Finally, let us describe  the inverse HP phase transition, which
is closely related to the evaporation process of the RBH.
 For $T=0.02>T_m$,  there is no meeting point between $F^{on}_{RBH}$
 and $F^{off}_{RBH}$ except $r_+=r_*$. For $T=T_m$, we find one
 meeting point (the minimum point) at $r_+=r_m$.
  For $T=0.005<T_m$, we find two meeting  points: unstable large black hole
at $r_+=r_i$ and extremal black hole at $r_+=r_*$.  Actually,
there is a change of dominance at the critical temperature
$T=0.005$: from unstable large black hole to stable extremal back
hole. Explicitly, the
 off-shell (non-equilibrium)  process starts with $r_+=r_i=15.72$ and ends at $r_+=r_e=1.84$.
We observe that $r_i \to \infty$ and $r_e \to r_*$, as $T \to 0$.
Hence this could be  regarded as the inverse HP transition for the
RBH.

{\it Evaporation of the regular black holes.---} We remind that
that the RBH looks like the RN black hole with the singularity
replaced by a regular center. The evaporating process will
terminate at the extremal point ($r_+=r_*$) before arriving at
$r_+=0$. Hence, as far as the evaporation process, there is no
difference between the regular and singular black holes. Following
Hayward~\cite{HAY} and Bonanno and Reuter~\cite{REU}, we find that
the early stage of evaporation is given by that of Schwarzschild
black hole.  The late stage of the evaporation process for the RBH
is totally different from the Schwarzschild case. Instead, this is
described by  the QCEP. We obtain the approximate forms for
temperature and luminosity:

\begin{eqnarray} && T_{RBH}(m) \simeq \alpha \sqrt{m-m_*},\label{3eq11} \\
    && L_{RBH}(m) \simeq \beta (m-m_*)^{2} \label{3eq12}
\end{eqnarray}
with $\alpha=3/8\pi m_*^2=0.07$ and $\beta=\sigma A
\alpha^4=9\sigma/64\pi^3m_*^6=0.00016$.  One finds
\begin{equation}
\label{3eq13} m(v)-m_{*} \propto \frac{1}{v},
\end{equation}
where $v$ is the advanced time coordinate. It was shown that
$m(v)-m_*$ vanishes as $v^{-1}$ for the RG-improved Vaidya
metric~\cite{REU}. Hence, we obtain the late stage  of the
evaporation process: $T_{\rm RBH}(v) \propto v^{-1}$ and $L_{RBH}
\propto v^{-4}$. We confirm that the RBHs lead to concrete
predictions on the final state of the evaporation process. We note
again that $m=m_*$ is the mass of a cold remnant, which is  an
extremal black hole with the Planck size. It takes an infinite time
to reach the extremal black hole, in compared with the Schwarzschild
black hole.

\begin{table}
\centering
\begin{tabular}{|c||c|c||c|c|}
\hline
 &\multicolumn{2}{c||}{HPT}&\multicolumn{2}{c|}{IHPT}\\
\cline{2-5}
 &starting point& ending point & starting point & ending point\\
\hline\hline
    $r_+$   & $r_+=0$ &$r_+=r_s$& $r_+=r_i$ &$r_+=r_\ast$\\ \hline
  $C_{RBH}$ &$0$ & $+$&$-$&0 \\ \hline
  $F_{RBH}$ &  0 & $-$  &  $+$ &0 \\ \hline
 stability  & TR & GSBH & UBH  & EBH \\
\hline
\end{tabular} \caption{Summary for the  Hawking-Page transition (HPT) and
 Inverse Hawking-Page phase transition (IHPT). In the bottom, TR (GSBH) means
 thermal radiation (globally
 stable black hole), and UBH (EBH) means
 unstable black hole (extremal black hole).}
\end{table}

{\it Discussions.---}
First of all, we mention that the local thermal stability is given
by positive heat capacity with $C>0$, while the global stability
is guaranteed for  positive heat capacity $C>0$ and the negative
free energy $F<0$.

We distinguish the difference between the Hawking-Page transition
in the Schwarzschild-AdS black hole and the inverse Hawking-Page
transition in the minimal model of the RBH.
The Hawking-Page transition is a thermodynamic process
by absorbing radiations in heat reservoir: thermal radiation
($C=0,F=0$) $\to$ unstable small black hole ($C<0,F>0$) $\to$
stable large black hole ($C>0,F<0$). Hence, the ending point is a
globally stable black hole. At the critical temperature $T=T_1$,
there is a change of the dominance from thermal radiation to a
black hole.

On the other hand, the inverse Hawking-Page transition is an
evaporation  process by emitting radiations through the Hawking
radiation: unstable large black hole ($C<0,F>0$)  $\to$ locally
stable black hole ($C>0,F>0$) $\to$ stable extremal black hole
($C=0,F=0$). The ending state is supposed to be a vacuum state
because it has $T=C=F=0$ and $S\not=0$. At the critical
temperature near $T=0$, there is a change of the dominance from
locally stable black hole to extremal black hole. These are
summarized in Table I.

Concerning the temperature $T$, which  defines the inverse
Hawking-Page transition, we have still some arguments for
regarding $T$ as the temperature of heat reservoir. This is
because we did not introduce any reservoir such as the cavity  for
the Schwarzschild black hole~\cite{York} and the negative
cosmological constant for the AdS black hole~\cite{Mann}. These
are necessary devices to derive the Hawking-Page transition from
thermal radiation to large stable black hole. Here, we have used the
external temperature $T$ by assuming the reservoir.

Moreover, in this work the backreaction effect is trivial because
the temperature approaches zero (not divergent) as $m \to m_*$. For
the Schwarzschild case, one expects relevant backreaction effects
during the terminal stage of the evaporation because of huge
increase of temperature as approaches $T=0$. However,  there is a
suppression of quantum backreaction for the RBH, since it emits less
and less energy as the QCEP does.

In summary, we have newly shown that the whole evaporation process
in the minimal model of the RBH could be regarded as the inverse
Hawking-Page phase transition comparing with the Hawking-Page phase
transition in the AdS black hole. Its early stage is described by
the evaporation of Schwarzschild black hole and its late stage is
described by the QCEP with $C>0$ and $F>0$.  In fact, our result is
universal for any RBHs although we have newly investigated the
thermodynamics and the evaporation process by choosing a minimal
model suggested by Hayward. This is because the temperature in Fig.
1, the heat capacity in Fig. 2, and the free energy in Fig. 3 show
the  same behaviors for all known RBHs including the loop quantum
and RN black holes.

\newpage
\section*{Acknowledgement}
Two of us (Y.S. Myung and Y.-J. Park) were supported by the Science
Research Center Program of the Korea Science and Engineering
Foundation through the Center for Quantum Spacetime of Sogang
University with grant number R11-2005-021.

\end{document}